\documentclass[a4paper]{jpconf}

\usepackage{amsmath,amssymb}
\usepackage{graphicx}

\newcommand{\mcb}{\mathcal{B}}
\newcommand{\mcl}{\mathcal{L}}
\newcommand{\mco}{\mathcal{O}}
\newcommand{\mcs}{\mathcal{S}}
\newcommand{\mcv}{\mathcal{V}}
\newcommand{\evbg}[1]{\left. #1 \right\vert_\text{bg}}

\begin{document}

\hfill{}NIKHEF/2010-035

\title{Stability of gravity-scalar systems for domain-wall models with a soft wall}
\author{Damien P. George}
\address{Nikhef Theory Group, Science Park 105, 1098 XG Amsterdam, The Netherlands}
\ead{dpgeorge@nikhef.nl}

\begin{abstract}
We show that it is possible to create an RS soft-wall model,
a model with a compact extra dimension, without using fundamental
branes.  All that is required are bulk scalar fields minimally
coupled to gravity.  Of crucial importance is the stability of
the size of the extra dimension.  Without branes, one cannot
easily implement the Goldberger-Wise mechanism, and instead it
must be shown that the scalar configuration is stable in its
own right.  We use the superpotential approach for generating
solutions, the so called 'fake supergravity' scenario, and show
that configurations generated in such a way are always free of
tachyonic modes.  Furthermore, we show that the model is also free
of zero modes (in the spin-0 sector) if all the scalars
have odd parity.  We discuss the hierarchy problem in soft-wall
models, and applications of our analysis to the AdS/QCD
correspondence.
\end{abstract}

\section{Introduction}
Extra dimensions are a plausible extension to the standard model
of particle physics.  In particular, much attention has been given
to the type I Randall-Sundrum (RS) model~\cite{Randall:1999ee},
where the electroweak hierarchy can be naturally generated by the
warping of a compact extra dimension.  In their original
realisation, these models have fundamental branes of negative
tension, or hard-walls, at the boundary of the extra dimension.
Recently there has been interest in a new type of RS-like warped
spacetime: a compact spacetime where the negative tension brane is
replaced with a physical singularity.  These soft-wall models were
originally designed to yield linear Regge trajectories in the
context of the AdS/CFT correspondence~\cite{Karch:2006pv}, but
have since been the basis of actual models beyond the standard
model~\cite{Batell:2008zm,Falkowski:2008fz,Batell:2008me,
Delgado:2009xb,MertAybat:2009mk,Gherghetta:2009qs,Cabrer:2009we,
vonGersdorff:2010ht}, and also provide a holographic dual
description of unparticle models~\cite{Cacciapaglia:2008ns,
Falkowski:2008yr}.

\begin{figure}
\begin{center}
\includegraphics[width=.95\textwidth]{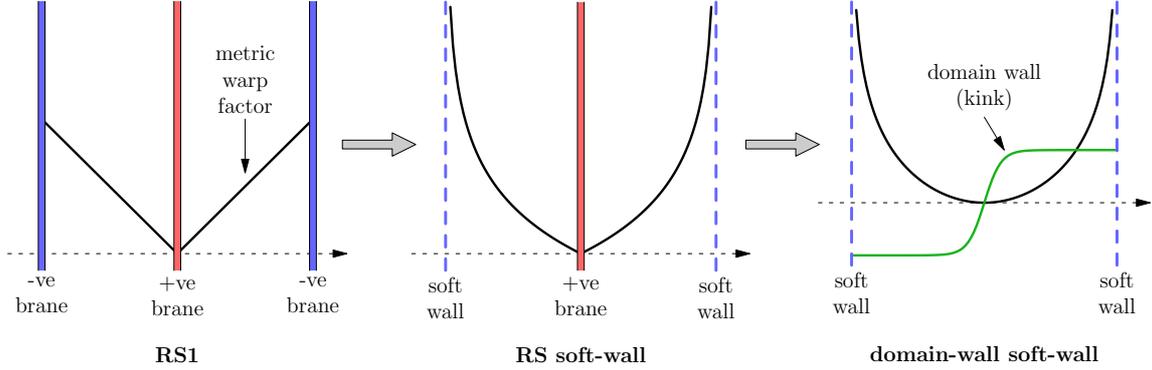}
\caption{From RS1 to RS soft-wall to domain-wall soft-wall.
The domain-wall set-up provides a purely field theoretic
construction of a compact extra dimension.
\label{fig:rs1todwsw}}
\end{center}
\end{figure}

Our aim is take the soft-wall model and go one step further by
removing the final, positive-tension brane at the origin, and
replacing it with a suitable scalar field profile, such as a
domain wall.  Such a model will then describe a compact extra
dimension without the use of fundamental dynamical branes.  This
is interesting because it is a purely field theoretical
construction and requires no appeal to string theory.
Figure~\ref{fig:rs1todwsw} depicts the progression from RS1
models to domain-wall soft-wall models.  In what follows, we
analyse the stability of such domain-wall models, and show that
they are capable of solve the hierarchy problem.  See
reference~\cite{Aybat:2010sn} for details.

\section{Warped Background Configuration}

We work in the framework of 5D general relativity, with $N$ scalar
fields $\Phi_i$ minimally coupled to gravity.  We also allow the
presence of brane terms for completeness.  The action for such a
set-up is given by
\begin{equation}
\mcs = \int d^4x \, dy
    \left[ \sqrt{-g} \left( M^3 R + \mcl_{\text{matter}} \right)
        - \sqrt{-g_4} \lambda
    \right] \:,
\end{equation}
where $M$ is the 5D Planck mass and the matter and brane terms are
\begin{align}
\mcl_\text{matter} &= -\frac{1}{2} \sum_{i=1}^N g^{PQ} \partial_P \Phi_i \partial_Q \Phi_i - V(\{\Phi_i\}) \:,\\
\lambda &= \lambda(\{\Phi_i\}) = \sum_\alpha \lambda_\alpha(\{\Phi_i\}) \delta(y-y_\alpha) \:.
\end{align}
Here, $V(\{\Phi_i\})$ is the scalar potential and
$\lambda_\alpha(\{\Phi_i\})$ the brane localised potential
(including brane tension) for the brane at $y_\alpha$.  Our aim is
to study the general stability conditions for non-trivial
background configurations of this class of models.

We specialise to backgrounds that depend only on the extra
dimension $y$, and denote the scalar background solutions
$\phi_i(y)$.  The RS background metric ansatz is
\begin{equation}
ds^2 = e^{-2\sigma(y)} \eta_{\mu\nu} dx^\mu dx^\nu + dy^2 \:,
\end{equation}
where $\mu,\nu$ index the 4D subspace.  Einstein's equations are
$G_{MN} = \frac{1}{2M^3} T_{MN}$ and yield, together with the
Euler-Lagrange equations,
\begin{align}
\label{eq:bg-1}
&6M^3 \sigma'' =
    \sum_i \phi_i'^2 + \lambda(\{\phi_j\}) \:, \\
\label{eq:bg-2}
&6M^3 \left( 4 \sigma'^2 - \sigma'' \right) =
    -2 V\left(\{\phi_j\}\right) - \lambda\left(\{\phi_j\}\right) \:, \\
\label{eq:bg-3}
&\phi_i'' - 4 \sigma' \phi_i'
    - V_i(\{\phi_j\})
    - \lambda_i(\{\phi_j\}) = 0 \:.
\end{align}
The notation $V(\{\phi_i\})$ means that the potential is to be
evaluated with the background fields $\phi_i$.  A subscript $i$ on
$V$ or $\lambda$ denotes partial differentiation with respect to
the field $\Phi_i$.

Given a particular $V$ and $\lambda$, the space of background
configurations is parameterised by the integration constants of
the system.  For a particular choice of integration constants,
that is, a particular background, we want to know if such a
choice is stable within the space of configurations.  We shall
study local, perturbative stability by adding small fluctuations
to the background, turning the problem into an eigenvalue
problem.  For the initial stages our analysis will be for an
arbitrary scalar potential $V$.  Later on we will need to
specialise to the fake supergravity approach.

\section{Perturbations}
The general ansatz which takes into account both spin-0 and spin-2
perturbations is
\begin{align}
&ds^2 = e^{-2 \sigma(y)} \left[ (1-2F(x^\mu,y) \eta_{\mu\nu} + h_{\mu\nu}(x^\mu,y) \right] dx^\mu dx^\nu
    + \left[1+G(x^\mu,y)\right]^2 dy^2 \:,\\
&\Phi_i(x^\mu,y) = \phi_i(y) + \varphi_i(x^\mu,y) \:.
\end{align}
Here we work in the axial gauge, $h_{\mu5}=0$, with transverse
traceless part $\partial^\mu h_{\mu\nu} = \eta^{\mu\nu} h_{\mu\nu}=0$.
The $(ij)$ Einstein's equations (off-diagonal spatial) enforce
$G=2F$ which we take from now on.  The rest of Einstein's
equations correspond to $(\mu\nu)$, $(\mu5)$ and $(55)$, and
taking the 4-trace of the $(\mu\nu)$ equations shows that the
spin-2 and spin-0 perturbations decouple from one another.

\subsection{Spin-2 Perturbations}

The spin-2 perturbation $h_{\mu\nu}$ decouples from $F$ and
$\varphi_i$.  The equation for $h_{\mu\nu}$ is
\begin{equation}
-e^{2\sigma} \Box h_{\mu\nu}
    - h_{\mu\nu}''
    + 4\sigma'h_{\mu\nu}' = 0 \:.
\end{equation}
There is always a zero mode, $h_{\mu\nu}'=0$, which is
normalisable.  This is the well-known 4D massless graviton from
the RS model~\cite{Randall:1999vf}. In conformal coordinates $z$
defined by $dy=e^{-\sigma}\,dz$ with rescaled
$h_{\mu\nu} = e^{3\sigma/2} \tilde{h}_{\mu\nu}$ we can write
this equation as a Schr\"odinger-like equation in a self-adjoint
form:
\begin{equation}
(\partial_z - \frac{3}{2}\sigma')(-\partial_z - \frac{3}{2}\sigma') \tilde{h}_{\mu\nu}
    = \Box \tilde{h}_{\mu\nu} \:.
\end{equation}
Applying supersymmetric quantum mechanics we find that there are
no tachyonic modes in the spin-2 sector.  Thus the spin-2
fluctuations do not destabilise the configuration.

\subsection{Spin-0 Perturbations}

The spin-0 sector is significantly more complicated than the
spin-2 sector, since physical modes are mixtures of $F$ and
$\varphi_i$.  The equations for these perturbations consist of two
of Einstein's equations and the Euler-Lagrange equations (sum
over repeated scalar indices):
\begin{align}
\label{eq:pert-flin-1}
&6M^3(F' - 2 \sigma' F) = \phi_i' \varphi_i \:,\\
\label{eq:pert-fquad-1}
&6M^3(-e^{2\sigma} \Box F - 2 \sigma' F' + F'') = 2 \phi_i' \varphi_i' + 2 \evbg{\lambda} F + \evbg{\lambda_i} \varphi_i \:.\\
\label{eq:pert-phi-1}
&e^{2\sigma} \Box \varphi_i + \varphi_i'' - 4 \sigma' \varphi_i' - 6 F' \phi_i'
    - \evbg{(4V_i + 2\lambda_i)} F
    - \evbg{(V_{ij} + \lambda_{ij})} \varphi_j = 0 \:.
\end{align}
To simplify these equations we go to conformal coordinates $z$ as
above and rescale the fields by
$F = e^{3\sigma/2} \chi(z)/\sqrt{12}$ and
$\varphi_i = M^{3/2} e^{3\sigma/2} \psi_i(z)$.
Equations~\eqref{eq:pert-flin-1} through~\eqref{eq:pert-phi-1}
then become a set of coupled Schr\"odinger-like equations:
\begin{equation}
\label{eq:sym-spin0}
\begin{aligned}
-\chi'' + (\mcv_{00} + \mcb_{00}) \chi + (\mcv_{0i} + \mcb_{0i}) \psi_i &=
    \Box \chi \:,\\
-\psi_i'' + (\mcv_{ij} + \mcb_{ij}) \psi_j + (\mcv_{0i} + \mcb_{0i}) \chi &=
    \Box \psi_i \:,
\end{aligned}
\end{equation}
where the effective potential $\mcv$ has components
\begin{align}
\mcv_{00} = \frac{9}{4} \sigma'^2 + \frac{5}{2} \sigma'' \:,\quad
\mcv_{0i} = \frac{2 \phi_i''}{\sqrt{3M^3}} \:,\quad
\mcv_{ij} = \left( \frac{9}{4} \sigma'^2 - \frac{3}{2} \sigma'' \right) \delta_{ij}
    + \frac{\phi_i'\phi_j'}{M^3} + e^{-2\sigma} \evbg{V_{ij}} \:,
\end{align}
and the brane terms are
\begin{align}
\mcb_{00} = \frac{1}{3M^3} e^{-2\sigma} \evbg{\lambda} \:,\qquad
\mcb_{0i} = \frac{1}{\sqrt{3M^3}} e^{-2\sigma} \evbg{\lambda_i} \:,\qquad
\mcb_{ij} = e^{-2\sigma} \evbg{\lambda_{ij}} \:.
\end{align}
Equation~\eqref{eq:sym-spin0} defines the physical spin-0
spectrum.  The symmetry of the cross-coupling ensures that the
eigenvalues of this spectrum are real.  To demonstrate stability
of a given background configuration, we must also show that all
eigenvalues are positive.

\section{Fake supergravity}

Our aim is to find a class of models which have strictly positive
eigenvalues in the spin-0 sector, ensuring the stability of the
background configuration.  We specialise to the case without brane
terms ($\lambda=0$) and where the scalar potential is generated
from a superpotential $W(\{\Phi_i\})$ using the fake supergravity
approach~\cite{DeWolfe:1999cp,Freedman:2003ax}
\begin{equation}
V(\{\Phi_i\})
= \sum_i \frac{1}{2} \left[ W_i(\{\Phi_i\}) \right]^2
- \frac{1}{3 M^3} \left[ W(\{\Phi_i\}) \right]^2 \:,
\end{equation}
where $W_{i}\equiv dW/d\Phi_i$.  For such a potential,
equations~\eqref{eq:bg-1} through~\eqref{eq:bg-3} simplify to
\begin{align}
\label{eq:bg-w}
\sigma'(y) = W(\{\phi_i\})/6M^3 \:,\qquad
\phi_i'(y) = W_i(\{\phi_i\})\:.
\end{align}
This is a set of \emph{first order} equations; $W$ encodes for
both $V$ and some of the integration constants.  Without loss of
generality we can take $\sigma(y_0)=0$, leaving $N$ integration
constants for the boundary values of $\phi_i$.  So, given a
particular $W$, the set $\{\phi_i(y_0)\}$ will then uniquely
define a configuration.  We want to know if such a configuration
is stable.

Using the fake supergravity approach, one can write the spin-0
effective potential as $\mcv = S^2 + S'$ where
\begin{equation}
\label{eq:s}
  S = e^{-\sigma}
  \left.
  \begin{pmatrix}
    \tfrac{1}{12 M^3} W &
    \,\,\tfrac{1}{\sqrt{3 M^3}} W_{j} \\
    \tfrac{1}{\sqrt{3 M^3}} W_{i} &
    \,\,\tfrac{-1}{4 M^3}\delta_{ij} W
      + W_{ij}
    \end{pmatrix}\right\vert_\text{bg} \:.
\end{equation}
Then equation~\eqref{eq:sym-spin0} takes the self-adjoint form
\begin{equation}
\label{eq:spin0-s}
(\partial_z + S^{\dagger})(-\partial_z + S)\Psi = \Box\Psi \:,
\end{equation}
where $\Psi=(\chi,\psi_i)$.  For a warped metric the relevant
boundary terms vanish and supersymmetric quantum mechanics tells
us that the eigenvalues of the spin-0 sector are non-negative.
The background is thus free of tachyonic modes.  To ensure
complete stability, we must still analyse the existence of zero
mode solutions.  Such zero modes can, for example, correspond to
changes in the size of the compact extra dimension, rendering the
set-up unstable.

\subsection{Zero modes with $N=1$}

With $N=1$ scalar we can use the Einstein constraint equation to
eliminate $\psi_1$ in terms of $\chi$.  Then let
$\chi=S_{01} g$, with $S_{01}$ defined by equation~\eqref{eq:s}.
The Schr\"odinger-like equation for $g$ is
\begin{equation}
  (\partial_z - S_{11})(-\partial_z - S_{11}) g + S_{01}^2 g = \Box g \:.
\end{equation}
Multiplying this equation from the left by $g^*$ and integrating
over the extra dimension yields
\begin{equation}
  \int \left\vert(-\partial_z - S_{11})g\right\vert^2 dz
    + \int \left\vert S_{01} g \right\vert^2
    + \text{(boundary terms)} = \Box \int\vert g \vert^2 dz\:.
\end{equation}
For a warped metric the boundary terms vanish, and the existence
of a zero mode with $\Box g=0$ requires both $(-\partial_z - S_{11})g=0$
and $S_{01}g=0$.  Thus $g=0$ and we can conclude that systems with
$N=1$ scalar using the fake supergravity approach do not have a
zero mode.

\subsection{Zero modes with $N\ge2$}

Set-ups with more than one scalar field can in general posses a
zero mode.  Nevertheless, we can formulate a simple criterion that
can be used to find models which do not have a zero mode: for a
system of definite parity, the number of independent normalisable
zero modes is at most equal to the number of even scalar fields.
This is essentially a statement about integration constants.  In
the fake supergravity approach, the background configuration is
given by the solutions to the first order equations~\eqref{eq:bg-w}.
The restriction to models with parity eliminates those integration
constants associated with odd-parity scalars, since their field
value must vanish at $y=0$.  Unique solutions to the fake
supergravity equations are then parameterised by the integration
constants of the even-parity fields.  The final point is that zero
modes move us continuously through this space of solutions, so
there cannot be more zero modes than the number of even fields.

\section{Applications}

We shall look at two example models with $N=2$ scalar fields to
illustrate the above criterion about the existence of zero modes.
The second model is an example of a stable domain-wall soft-wall
set-up.  Following this we make some comments about AdS/QCD.

\subsection{Example 1 --- unstable even system}

\begin{figure}
  \begin{minipage}[t]{0.48\textwidth}
    \includegraphics[width=\textwidth]{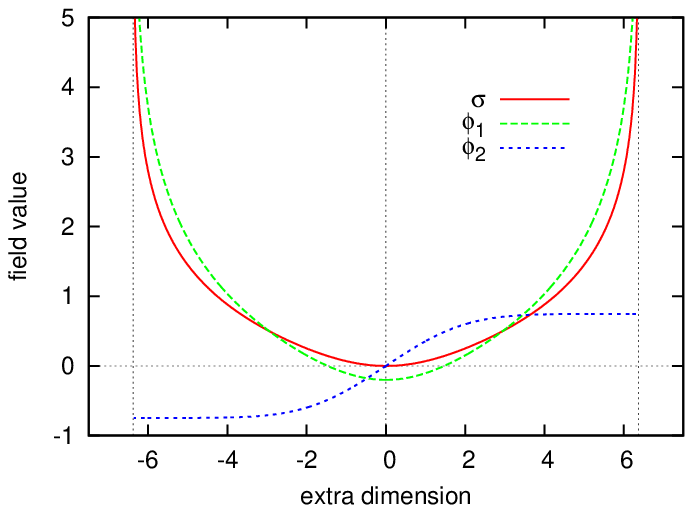}
    \caption{
    Background solutions for Example 1.
    Parameters are $\nu=1.4$, $a=0.5$, $b=0.3$.
    The size of the extra dimension is not stabilised and depends
    on the value $\phi_1(0)$.
    \label{fig:ex1}}
  \end{minipage}
  \hfill
  \begin{minipage}[t]{0.48\textwidth}
    \includegraphics[width=\textwidth]{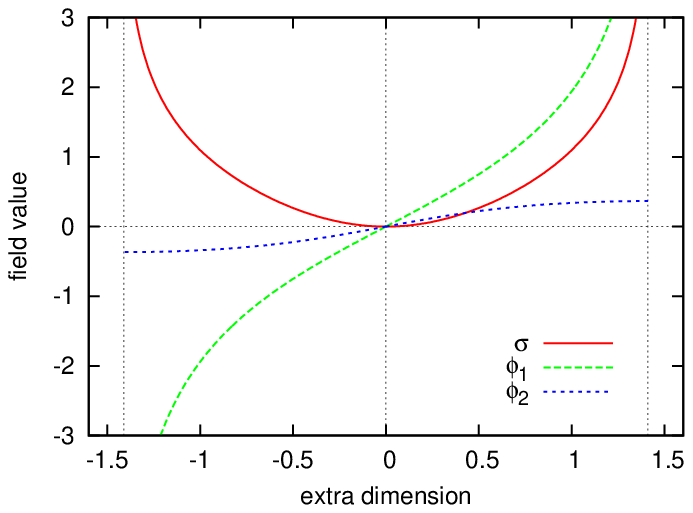}
    \caption{
    Background solutions for Example 2.
    Parameters are $\alpha=1$, $\nu=1.4$, $a=0.5$, $b=0.3$.
    The configuration is stable and is an example of a
    domain-wall soft-wall model.
    \label{fig:ex2}}
  \end{minipage}
\end{figure}

Consider the superpotential
\begin{equation}
W\left(\Phi_1,\Phi_2\right)
    = e^{\nu\Phi_1} \left(a\,\Phi_2 - b\,\Phi_2^3\right) \:,
\end{equation}
where $\Phi_1$ and $\Phi_2$ are the dilaton and kink fields
respectively.  We choose $\Phi_1$ to have even and $\Phi_2$ to
have odd parity in $y$.  See Figure~\ref{fig:ex1} for
representative background solutions (the plots use units where
$6M^3=1$).  In these solutions there is
a choice of integration constant for $\Phi_1$ and our criterion
tells us that a corresponding spin-0 zero mode can exist.  Such
a mode in fact does exist, and is given by
$(\chi,\psi_1,\psi_2)=N(-\nu,\sqrt{2},0)e^{-3\sigma/2}\eta(x^\mu)$
where $N$ is a normalisation constant and $\eta$ is the 4D mode.
This zero mode physically corresponds to changes in the size of
the extra dimension, so the background configuration is not
stable.

\subsection{Example 2 --- stable odd system}

\begin{figure}
  \begin{center}
    \includegraphics[width=.5\textwidth]{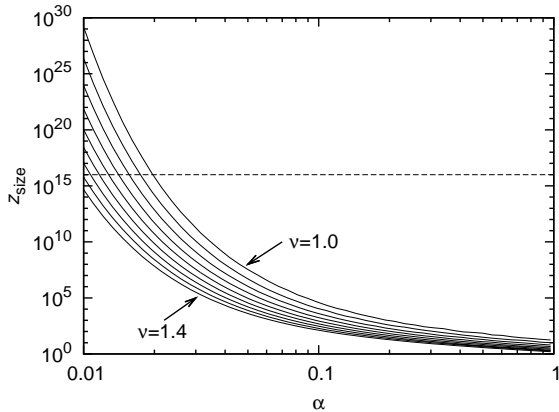}
    \hfill
    \begin{minipage}[b]{0.46\textwidth}
      \caption{Solving the hierarchy problem using the domain-wall
      soft-wall model of Example 2.
      The size of the compact extra dimension in conformal
      coordinates, $z_\text{size}$, sets the characteristic
      mass scale.  Bulk fields have a KK mass scale
      $m_\text{KK}\sim z_\text{size}^{-1}$ and the hierarchy
      problem is solved if
      $z_\text{size} \sim 10^{16}$ can be obtained for $\mco(1)$
      model parameters.  The plot has $a=b=1$ in
      equation~\eqref{eq:ex2w} and shows $z_\text{size}$ versus
      $\alpha$ for $\nu=1.0\ldots1.4$.
      To obtain the relevant hierarchy one should take
      $\alpha\simeq0.02$ and $\nu\simeq1.0$.
      \label{fig:hierarchy}}
    \end{minipage}
  \end{center}
\end{figure}

The second model has the superpotential
\begin{equation}
\label{eq:ex2w}
  W\left(\Phi_1,\Phi_2\right)
    = \alpha\,\sinh(\nu\,\Phi_1) + \left(a\,\Phi_2 - b\,\Phi_2^3\right) \:,
\end{equation}
where $\Phi_1$ and $\Phi_2$ are now both taken to have odd parity.
As a consequence of this parity choice, there are no integration
constants to choose, and, by our criterion, there cannot exist any
zero modes.  The background solution is unique and stabilised, and
the size of the extra dimension is fixed by the parameters in $W$;
see Figure~\ref{fig:ex2}.  The field $\Phi_2$ is in a kink
configuration and replaces the brane in usual RS soft-wall models,
providing an example of a domain-wall model with a soft wall.
Figure~\ref{fig:hierarchy} shows how the electroweak hierarchy can
be generated in such a model.

\subsection{Application to AdS/QCD}

Soft-wall models were originally motivated by attempts to
construct 5D AdS models of QCD~\cite{Karch:2006pv}.  Having a soft
wall in the infrared (large $y$) yields linear Regge trajectories;
the meson excitations in the QCD dual theory have masses which
scale like $m_n^2 \sim n$, where $n$ is the excitation number.
Scalar fluctuations in the AdS picture correspond to glueball and
scalar meson excitations in the dual QCD theory.  Since the fake
supergravity approach is common in the literature, and it seems
that more than one scalar is required to dynamically generate
the background (see for example~\cite{Batell:2008zm}), our
Schr\"odinger-like equation for the spin-0 eigenvalues,
equations~\eqref{eq:s} and~\eqref{eq:spin0-s}, can be used to
compute the scalar spectrum with multiple background fields.

\enlargethispage{2pt}

\ack
This research was done in collaboration with S.~M.~Aybat.
It was supported by the Netherlands Foundation for
Fundamental Research of Matter (FOM) and the National Organisation
for Scientific Research (NWO).

\section*{References}

\bibliographystyle{iopart-num}
\bibliography{references}

\end{document}